\documentstyle[pra,aps]{revtex} 
\begin{document} 
  \draft 
  \title{GEOMETRICAL ORIGIN OF FERMION FAMILIES
\\
 IN $SU(2)\times U(1)$ GAUGE THEORY}
\author{E. I. Guendelman \thanks{guendel@bgumail.bgu.ac.il} and
A.  B.  Kaganovich \thanks{alexk@bgumail.bgu.ac.il}} 
\address{Physics Department, Ben Gurion University of the Negev, Beer Sheva 
84105, Israel}
\maketitle
\begin{abstract} 
A spontaneously broken $SU(2)\times U(1)$ gauge theory with just one
"primordial" generation of fermions is formulated in the context of
generally covariant theory which contains two measures of integration
in the action: the standard $\sqrt{-g}d^{4}x$ and a new $\Phi d^{4}x$,
where $\Phi$ is a density built out of degrees of freedom independent 
of the metric. Such type of models are known to produce a satisfactory 
answer  to the cosmological constant problem. Global scale invariance 
is implemented. After SSB of scale invariance and gauge symmetry
it is found that with the conditions appropriate to laboratory particle
physics experiments, to each 
primordial fermion field corresponds three physical fermionic states.
Two of them correspond to particles with different constant masses and
they are 
identified with the first two generations of the electro-weak theory.
The third fermionic states  at the classical level get non-polynomial
interactions which indicate the existence of fermionic condensate
and fermionic mass generation.

   \end{abstract}          
    
    \renewcommand{\baselinestretch}{1.6} 

\pagebreak 

{\bf I. INTRODUCTION.} 
One of the most perplexing questions that have arisen in the theory of
elementary particles is the origin of the families (generations) of
elementary fermions: electrons and quarks. Indeed,   each fermion is
replicated three times: instead of having one electron, we observe in
addition the muon and the tau lepton; instead of one quark doublet we have
three doublets of quarks. All these replications exhibit the same charge,
spin, etc. but they differ in their masses.

Numerous geometrical attempts to understand the origin of three fermion
families have been pursued in the context of string theory (for a
review see \cite{Zurab}).   
In this paper we will follow a different geometric approach to the family
problem of
particle physics. Basic ideas and methods of this approach have been
developed in previous papers\cite{GK1,GK2,GK3,G,K,GK5} where the emphasis
was
on cosmological
questions, in special the question of the cosmological constant problem.
It was noticed however\cite{GK5} that a natural solution to the family
problem could be  given along these lines as well. Here we generalize the
results of the toy model\cite{GK5} to the $SU(2)\times U(1)$ gauge theory. 

The geometric approach of Refs.\cite{GK1,GK2,GK3,G,K,GK5}
 consists of using an alternative
volume element $\Phi d^{4}x$, in addition to the standard one
$\sqrt{-g}d^{4}x$. So a general action of the form
\begin{equation}
    S = \int L_{1}\Phi d^{4}x +\int L_{2}\sqrt{-g}d^{4}x
\label{S}
\end{equation}
is considered. In order that $\Phi d^{4}x$ be an invariant volume
element, it is necessary that $\Phi$ transforms as a density, i.e. just
like $\sqrt{-g}$. This can be realized if we choose $\Phi$ to be the
composite of 4 scalars $\varphi_{a}$ ($a=1,2,3,4$)
\begin{equation}
\Phi
=\varepsilon^{\mu\nu\alpha\beta}\varepsilon_{abcd}\partial_{\mu}\varphi_{a}
\partial_{\nu}\varphi_{b}\partial_{\alpha}\varphi_{c}
\partial_{\beta}\varphi_{d}.
\label{Phi}
\end{equation}
Since $\Phi$ is a total derivative, a shift of $L_{1}$ by a
constant, $L_{1}\rightarrow L_{1}+const$, has the effect of adding to S
the integral of a total derivative , which does not change equations of
motion.
This is why the 
introduction of a new volume element has consequences on the way we think
about the cosmological constant problem\cite{GK2,GK3}.

In Eq. (\ref{S}), $L_{1}$ and $L_{2}$ are Lagrangian which are functions
of the matter
fields, the metric, the connection  (or spin-connection ) but not of the
"measure fields" $\varphi_{a}$. In such a case the action (\ref{S}) has
the infinite dimensional symmetry\cite{GK3}:
$\varphi_{a}\rightarrow\varphi_{a}+f_{a}(L_{1})$, where $f_{a}(L_{1})$
is an arbitrary function of  $L_{1}$.

It may appear at first sight  strange to think that geometry (measure,
connections, metric) are relevant to particle physics. This is because we
are used to think that these geometrical objects can be only related to
gravity. However, as we will see, the consistency condition of equations
of motion determines the ratio of two measures
\begin{equation}
\zeta \equiv\frac{\Phi}{\sqrt{-g}}
\label{zeta}
\end{equation}
as a function of matter fields. The surprising feature of the theory is
 that neither Newton constant nor curvature appears in this constraint
which means that the {\it geometrical scalar field} $\zeta (x)$
is determined by the matter fields configuration  locally and 
straightforward (that is without gravitational interaction). As we will
see, $\zeta (x)$ has a decisive influence in the determination of
particle masses and in the "families birth effect".
Therefore "Geometry" will be of importance, beyond
what was  known so far, i. e. that the geometrical objects which 
enter into the field theory are  restricted by the metric associated to
the 
gravitational field and possibly  torsion  and  non-metricity\cite{Neem}.

{\bf II. THE MODEL.}
To see how the theory works, let us consider a model containing the
$SU(2)\times U(1)$ gauge structure (the color SU(3) can be added without 
changing our results), as in the standard model with sdandard content
of the bosonic sector (gauge vector fields $\vec{A}_{\mu}$ and $B_{\mu}$
 and Higgs doublet $H$). But in contrast to the standard model,
in our model {\it we start from only one family of the so called "primordial"
fermionic fields}:  the primordial up and down quarks $U$ and $D$  and
 the primordial electron $E$ and neutrino  $N$. Similar to the standard
model, we
will proceed with the following independent fermionic degrees of freedom:

a) one primordial left quark $SU(2)$ doublet $Q_{L}$ 
\begin{displaymath}
Q_{L}=\left( \begin{array}{c}
U_{L} \\
D_{L}
\end{array} \right)
\label{QL}
\end{displaymath}
and right primordial singlets $U_{R}$ and $D_{R}$;\\

b) one primordial left lepton SU(2) doublet $L_{L}$:
\begin{displaymath}
L_{L}=\left( \begin{array}{c}
N_{L} \\
E_{L}
\end{array} \right)
\label{LL}
\end{displaymath}
and right primordial singlet $E_{R}$.

In addition, a dilaton
field $\phi$ is needed in order to achieve global scale 
invariance\cite{G}.

According to the general prescriptions of the two measures theory,
we have to start from studying the selfconsistent system of gravity and
matter fields proceeding in the first order formalism. In the model including
fermions in curved space-time, this means that the
independent dynamical degrees of freedom are: all  matter fields,  
 vierbein $e_{a}^{\mu}$, spin-connection $\omega_{\mu}^{ab}$ and the measure
$\Phi$ degrees of freedom, i.e. four scalar fields $\varphi_{a}$. 
We postulate that in  addition to $SU(2)\times U(1)$ gauge symmetry,
the theory is invariant under the global scale transformations:
\begin{eqnarray}
    e_{\mu}^{a}\rightarrow e^{\theta /2}e_{\mu}^{a}, \quad
\omega^{\mu}_{ab}\rightarrow \omega^{\mu}_{ab}, \quad
\varphi_{a}\rightarrow \lambda_{a}\varphi_{a}\quad
where \quad \Pi\lambda_{a}=e^{2\theta}, 
\nonumber
\\ 
\phi\rightarrow \phi-\frac{M_{p}}{\alpha}\theta ,\quad
H\rightarrow H , \quad
\Psi\rightarrow e^{-\theta /4}\Psi, \quad 
\overline{\Psi}\rightarrow  e^{-\theta /4} \overline{\Psi};
\quad \theta =const. 
\label{stferm} 
\end{eqnarray}
This global scale invariance is important for cosmological applications
of the theory\cite{G,K,GK5}.

The action of the model has the general structure given by Eq. (\ref{S})
which is convenient to represent in the following form:
\begin{eqnarray}
&S=&\int d^{4}x e^{\alpha\phi /M_{p}}
(\Phi +b\sqrt{-g})\left[-\frac{1}{\kappa}R(\omega ,e) 
+\frac{1}{2}g^{\mu\nu}\phi_{,\mu}\phi_{,\nu}+
\frac{1}{2}g^{\mu\nu}(D_{\mu}H)^{\dag}D_{\nu}H\right]
-\int d^{4}x e^{2\alpha\phi /M_{p}}[\Phi V_{1}(H) +\sqrt{-g}
V_{2}(H)]
\nonumber\\
&&-\int d^{4}x\sqrt{-g}
\left(\frac{1}{4}g^{\alpha\beta}g^{\mu\nu}B_{\alpha\mu}B_{\beta\nu}
+\frac{1}{2}g^{\alpha\beta}g^{\mu\nu}Tr\,A_{\alpha\mu}A_{\beta\nu}\right) 
     +\int d^{4}x e^{\alpha\phi /M_{p}}(\Phi +k\sqrt{-g})L_{fk}
\nonumber\\
&&-\int d^{4}xe^{\frac{3}{2}\alpha\phi /M_{p}}
\left[(\Phi
+h_{E}\sqrt{-g})
       f_{E}\overline{L_{L}}\,H\,E_{R}
+(\Phi +h_{U}\sqrt{-g})
       f_{U}\overline{Q_{L}}\,\tilde{H}\,U_{R} 
+(\Phi +h_{D}\sqrt{-g})
       f_{D}\overline{Q_{L}}\,H\,D_{R} +H.c.\right]
\label{totaction}
\end{eqnarray}

The notations in (\ref{totaction}) are the following: 
$g^{\mu\nu}=e^{\mu}_{a}e^{\nu}_{b}\eta^{ab}$; the scalar curvature is
$R(\omega ,V) =V^{a\mu}V^{b\nu}R_{\mu\nu ab}(\omega)$ where
\begin{equation} 
R_{\mu\nu ab}(\omega)=\partial_{\mu}\omega_{\nu ab}
+\omega_{\mu a}^{c}\omega_{\nu cb}
-(\mu\leftrightarrow\nu);
        \label{B}
\end{equation}
 \begin{equation}
D_{\mu}H\equiv
\left(\partial_{\mu}
-\frac{i}{2}g\vec{\tau}\cdot\vec{A}_{\mu}
-\frac{i}{2}\tilde{g^{\prime}}B_{\mu}\right)H;
\label{DHiggs}
 \end{equation}
\begin{equation}
L_{fk}=\frac{i}{2}\left[\overline{L_{L}}\not\!\!D^{(L)}L_{L}+
  \,\overline{E}_{R}\not\!\!D^{(R)}E_{R}+
  \,\overline{Q}_{L}\not\!\!D^{(L)}Q_{L}+
  \,\overline{U}_{R}\not\!\!D^{(R)}U_{R}+
  \,\overline{D}_{R}\not\!\!D^{(R)}D_{R}\right]
\label{Lfkin}
\end{equation}
where
\begin{equation}
D^{(L)}\equiv\overrightarrow{\not\!\!D}_{L}-\overleftarrow{\not\!\!D}_{L};
\qquad 
D^{(R)}\equiv\overrightarrow
{\not\!\!D}_{R}-\overleftarrow{\not\!\!D}_{R};
\label{DEL}
 \end{equation}
\begin{equation}
\overrightarrow{\not\!\!D}_{L}\equiv
e^{\mu}_{a}\gamma^{a}\left(\vec{\partial}_{\mu}
+\frac{1}{2}\omega_{\mu}^{cd}\sigma_{cd}I-
\frac{i}{2}g\vec{\tau}\cdot\vec{A}_{\mu}
+\frac{i}{2}g^{\prime}B_{\mu}\right); \quad
\overleftarrow{\not\!\!D}_{L}\equiv
\left(\overleftarrow{\partial}_{\mu}
-\frac{1}{2}\omega_{\mu}^{cd}\sigma_{cd}I+
\frac{i}{2}\,g\vec{\tau}\cdot\vec{A}_{\mu}
-\frac{i}{2}g^{\prime}B_{\mu}\right)\gamma^{a}e_{a}^{\mu}
\label{D12L}
 \end{equation}
where $I$ is $2\times 2$ unit matrix in the isospin space;
\begin{equation}
\overrightarrow{\not\!\!D}_{R}\equiv e^{\mu}_{a}\gamma^{a}
\left(\vec{\partial}_{\mu}
+\frac{1}{2}\omega_{\mu}^{cd}\sigma_{cd}
+ig^{\prime}B_{\mu}\right); \quad
\overleftarrow{\not\!\!D}_{R}\equiv
\left(\overleftarrow{\partial}_{\mu}
-\frac{1}{2}\omega_{\mu}^{cd}\sigma_{cd}
-ig^{\prime}B_{\mu}\right)\gamma^{a}e_{a}^{\mu};
\label{DR2}
 \end{equation}
and finally $B_{\mu\nu}\equiv\partial_{\mu}B_{\nu}-
                               \partial_{\nu}B_{\mu}$,  
$A_{\mu\nu}\equiv\partial_{\mu}A_{\nu}-
                               \partial_{\nu}A_{\mu}-
ig[A_{\mu}A_{\nu}-A_{\nu}A_{\mu}]$
where $A_{\mu}=\frac{1}{2}\vec{A}_{\mu}\cdot\vec{\tau}$.

A few explanations concerning our choice of the action (\ref{totaction})
are in necessary:

1) In order to avoid a possibility of negative energy contribution
from the space-time derivatives of the dilaton $\phi$ and Higgs
$H$ fields (see Ref.\cite{GK5}) we have chosen the
coefficient $b$ in front of $\sqrt{-g}$ in the first
integral of (\ref{totaction}) to
be a common factor of the gravitational term
 $-\frac{1}{\kappa}R(\omega ,e)$ and of the kinetic terms for $\phi$
 and $H$. This guarantees that this item can not be an
origin of ghosts in quantum theory. 

2) For the same reasons we choose
the kinetic terms of the gauge bosons in the conformal invariant   
form which is possible only if these terms are coupled to the
measure $\sqrt{-g}$. Introducing the coupling of these terms to
the measure $\Phi$
would lead to the nonlinear equations and non positivity of the
energy.

3) For simplicity, we have taken the coupling
of the kinetic terms of the fermions to the measures to be universal
(see the forth integral in Eq.(\ref{totaction})). 

Except for these three items, Eq.(\ref{totaction}) describes the most
general action
of the two measures theory satisfying the formulated above symmetries.
For example, one can introduce two different Higgs
potentials $V_{1}(H)$ and $V_{2}(H)$ coupled to the measures   
$\Phi$ and $\sqrt{-g}$
respectively. For the same reason there are two different sets of
Yukawa coupling of the fermions to Higgs field. 
Such general structure of the action has a crucial role since as we will
see, it provides a very specific nonlinear equation (constraint) determining
the scalar
field
$\zeta$, Eq. (\ref{zeta}).
The multiple solutions of the nonlinear equation will be shown to
be associated to the "families birth effect".

{\bf III. CLASSICAL EQUATIONS OF  MOTION. }
After SSB of scale and gauge symmetries, proceeding in
the unitary gauge,
the Higgs field can be represented in the standard form
\begin{displaymath}
H=\left( \begin{array}{c}
0 \\
\frac{1}{\sqrt{2}}(\upsilon +\chi)
\end{array} \right)
\label{VEV}
\end{displaymath}
We will see later how in this theory the vacuum expectation value (VEV) 
$\upsilon$ is determined by an effective potential which is a very special
function of $V_{1}({H})$, $V_{2}({H})$ and the dilaton $\phi$.
 
Varying the measure fields $\varphi_{a}$, we get
\begin{equation} 
A^{\mu}_{a}\partial_{\mu}L_{1}=0 
\label{dl0}
\end{equation}
where $L_{1}$ is defined, according to  Eq. (\ref{S}), as
the part of the integrand of the action (\ref{totaction})
coupled to the measure $\Phi$ and
\begin{equation} 
A^{\mu}_{a}=\varepsilon^{\mu\nu\alpha\beta}\varepsilon_{abcd}
\partial_{\nu}\varphi_{b}\partial_{\alpha}\varphi_{c}
\partial_{\beta}\varphi_{d}.
\label{Ama}
\end{equation}

Since
$Det (A^{\mu}_{a})
= \frac{4^{-4}}{4!}\Phi^{3}$ it follows that if $\Phi\neq 0$, 
\begin{equation}
 L_{1}=sM^{4} =const
\label{varphi}
\end{equation}
where $s=\pm 1$ and $M$ is a constant of integration of the dimension of
mass.
It can be noticed that the appearance of a nonzero integration 
constant $M^4$ spontaneously breaks the scale invariance (\ref{stferm}).

Complete system of equations corresponding to the action
(\ref{totaction}) is very bulky. Variation of $S$
with respect to vierbein $e^{\mu}_{a}$ yields
the gravitational equation linear both in the curvature and in the
scalar field $\zeta$, defined by Eq. (\ref{zeta}). Contracting this
equation with $e^{\mu}_{a}$, solving for the curvature scalar $R$
and replacing in Eq. (\ref{varphi}) we obtain the following consistency
condition of the theory:
\begin{eqnarray} 
&&(\zeta -b)\left[sM^{4}e^{-\alpha\phi /M_{p}}+
               V_{1}(\upsilon +\chi)e^{\alpha\phi /M_{p}}-L_{fk}
     +\frac{1}{\sqrt{2}}(\upsilon +\chi)e^{\frac{1}{2}\alpha\phi /M_{p}}
        (f_{E}\overline{E}E + f_{U}\overline{U}U+f_{D}\overline{D}D)\right]
\nonumber\\
&&+2V_{2}(\upsilon +\chi)e^{\alpha\phi
/M_{p}}+\frac{1}{2}(\zeta-3k)L_{fk}
+\sqrt{2}(\upsilon +\chi)e^{\frac{1}{2}\alpha\phi /M_{p}}
        (f_{E}h_{E}\overline{E}E + f_{U}h_{U}\overline{U}U+
f_{D}h_{D}\overline{D}D)=0
\label{constraintorig}
\end{eqnarray}
where $L_{fk}$ is defined by Eq. (\ref{Lfkin}).
Making use of equations  of motion for all the fermionic fields,  it is
easy
to check that the following  relation is true
\begin{equation}
L_{fk}=\frac{ e^{\frac{1}{2}\alpha\phi /M_{p}}}{\zeta +k}
\left[(\zeta +h_{E})
       f_{E}\overline{L_{L}}\,H\,e_{R}
+(\zeta +h_{U})
       f_{U}\overline{Q_{L}}\,\tilde{H}\,U_{R}
+(\zeta +h_{D})
       f_{D}\overline{Q_{L}}\,H\,D_{R} +H.c.\right]
\label{Lfkyuk}
  \end{equation}
Due to this relation, the consistency condition
(\ref{constraintorig}) becomes a constraint 
having a fundamental role for the theory. 

In order to get the physical content of the theory it is required
to express it in terms of variables where all equations of motion
acquire a canonical form in an Einstein-Cartan space-time (for
detail see Ref.\cite{GK3}). This is possible after performing the 
following redefinitions of the vierbein (and metric) and all fermion
fields (we are using here $\Psi$ as a common notation for all primordial
fermions):
\begin{equation}
\tilde{g}_{\mu\nu}=e^{\alpha\phi/M_{p}}(\zeta +b)g_{\mu\nu}, \quad  
\tilde{e}_{a\mu}=e^{\frac{1}{2}\alpha\phi/M_{p}}(\zeta
+b)^{1/2}e_{a\mu}, \quad
\Psi^{\prime}=e^{-\frac{1}{4}\alpha\phi/M_{p}}
\frac{(\zeta +k)^{1/2}}{(\zeta +b)^{3/4}}\Psi .
\label{ctferm}
\end{equation}

With these variables, the spin-connections become those of the 
Einstein-Cartan space-time. Since  $\tilde{e}_{a\mu}$ and
$\Psi^{\prime}$ are invariant under the scale transformations
(\ref{stferm}), spontaneous breaking of the scale symmetry (\ref{stferm})
(by means of Eq. (\ref{varphi}))  is reduced in the new variables to the
spontaneous breaking of the shift symmetry 
$\phi\rightarrow\phi +const$ for the dilaton field.

One can check  that equations of motion for the gauge fields in the new
variables are canonical and after the Higgs develops VEV, the gauge bosons
mass generation is standard,
that is exactly the same as it is in the Weinberg-Salam electroweak
theory: photon, $W^{\pm}$ and $Z$ bosons as well as the Weinberg
angle appear as the result of the standard procedure of
the Weinberg-Salam theory.

The gravitational equations of motion in the new variables take the form
\begin{equation}
G_{\mu\nu}(\tilde{g}_{\alpha\beta})=\frac{\kappa}{2}T_{\mu\nu}^{eff}
 \label{gef}
\end{equation}
\begin{eqnarray}
T_{\mu\nu}^{eff}& = & \phi_{,\mu}\phi_{,\nu}-K_{\phi}\tilde{g}_{\mu\nu}
+\chi_{,\mu}\chi_{,\nu}-K_{\chi}\tilde{g}_{\mu\nu}
+\tilde{g}_{\mu\nu}V_{eff}
+T_{\mu\nu}^{(gauge,can)}+T_{\mu\nu}^{(ferm,can)}
\nonumber\\
&&-\tilde{g}_{\mu\nu}[F_{E}(\zeta ,\upsilon +\chi)
\overline{E^{\prime}}E^{\prime}
+ F_{U}(\zeta ,\upsilon +\chi)\overline{U^{\prime}}U^{\prime}
+F_{D}(\zeta ,\upsilon +\chi)\overline{D^{\prime}}D^{\prime}],
 \label{Tmn}
\end{eqnarray}
Here $G_{\mu\nu}(\tilde{g}_{\alpha\beta})$ is the Einstein tensor
in the Riemannian (or,  more exactly,
 Einstein-Cartan) space-time with
metric
$\tilde{g}_{\mu\nu}$; $K_{\phi}\equiv\frac{1}{2}
\tilde{g}^{\alpha\beta}\phi_{,\alpha}\phi_{,\beta}$; 
$K_{\chi}\equiv\frac{1}{2}
\tilde{g}^{\alpha\beta}\chi_{,\alpha}\chi_{,\beta}$;
$T_{\mu\nu}^{(gauge,can)}$ is the canonical energy momentum tensor for
 gauge bosons, including mass terms of $W^{\pm}$ and $Z$ bosons.
  $T_{\mu\nu}^{(ferm,can)}$ is the canonical energy momentum tensor for
(primordial) fermions $E^{\prime}$, $U^{\prime}$ and $D^{\prime}$ in
curved space-time\cite{Birrel} 
including also their standard electromagnetic and weak interactions
 with gauge bosons.
Functions $V_{eff}$ and
$F_{i}(\zeta ,\upsilon +\chi)$ ($i=E^{\prime},U^{\prime},D^{\prime}$) are
defined by equations
\begin{equation}
V_{eff}=
\frac{b\left(sM^{4}e^{-2\alpha\phi/M_{p}}+V_{1}\right)-V_{2}}{(\zeta
+b)^{2}}
\label{Veff}
\end{equation}
\begin{equation}
F_{i}(\zeta ,\upsilon +\chi)\equiv
\frac{(\upsilon +\chi)f_{i}}{2\sqrt{2}(\zeta +k)^{2}(\zeta +b)^{1/2}}
[\zeta^{2}+(3h_{i}-k)\zeta +2b(h_{i}-k)+kh_{i}], \qquad 
i=E^{\prime},U^{\prime},D^{\prime}.
 \label{Fizeta}
\end{equation}

The scalar field $\zeta$ in the above equations is defined
by the constraint determined by means of Eqs. (\ref{constraintorig})
and (\ref{Lfkyuk}). In the new variables
(\ref{ctferm}) this constraint takes the form
\begin{equation}
(\zeta -b)\left[sM^{4}e^{-2\alpha\phi/M_{p}}+V_{1}(\upsilon +\chi)\right]+
2V_{2}(\upsilon +\chi)
+(\zeta +b)^{2}[F_{E}\overline{E^{\prime}}E^{\prime} +
F_{U}\overline{U^{\prime}}U^{\prime}
+F_{D}\overline{D^{\prime}}D^{\prime}
]=0.
\label{constraint}
\end{equation}

The dilaton $\phi$ and Higgs $\chi$ field equations in the new variables
are the following
\begin{equation}
\Box\phi -\frac{\alpha}{M_{p}(\zeta +b)} 
\left[sM^{4}e^{-2\alpha\phi/M_{p}}-\frac{(\zeta -b)V_{1}+2V_{2}}{\zeta
+b}\right]=
-\frac{\alpha }{M_{p}}[F_{E}\overline{E^{\prime}}E^{\prime} +
F_{U}\overline{U^{\prime}}U^{\prime}
+F_{D}\overline{D^{\prime}}D^{\prime}],
\label{phief+ferm}
\end{equation}
\begin{equation}
\Box\chi +\frac{\zeta V_{1}^{\prime}+V_{2}^{\prime}}{(\zeta +b)^{2}}=
-\frac{1}{(\zeta
+b)^{1/2}(\zeta +k)}[f_{E}(\zeta +h_{E})\overline{E^{\prime}}E^{\prime}
+ f_{U}(\zeta +h_{U})\overline{U^{\prime}}U^{\prime}
+f_{D}(\zeta +h_{D})\overline{D^{\prime}}D^{\prime}],
\label{chief+ferm}
\end{equation}   
where $\Box\phi =(-\tilde{g})^{-1/2}\partial_{\mu}
(\sqrt{-\tilde{g}}\tilde{g}^{\mu\nu}\partial_{\nu}\phi)$ and similarly for
$\Box\chi$.

Equations for the primordial fermions $E^{\prime}$, $U^{\prime}$
and $D^{\prime}$
in terms of the
variables (\ref{ctferm}) take the standard form of fermionic equations
in the Einstein-Cartan space-time\cite{Hehl} where the standard
interactions
 to the gauge fields present also. All the novelty consists of the form of
the $\zeta$ depending "masses" $m_{i}(\zeta)$ of the primordial fermions:
\begin{equation}
m_{i}(\zeta)=
\frac{f_{i}\upsilon (\zeta +h_{i})}{\sqrt{2}(\zeta +k)(\zeta +b)^{1/2}}
\qquad i=E^{\prime},U^{\prime},D^{\prime}.
 \label{muferm} 
\end{equation} 

{\bf IV. VACUUM AND FAMILIES BIRTH EFFECT.} Let us consider  the following
two
limiting cases:

(i) {\bf In the absence of massive fermions}, solving $\zeta$ from the
constraint
(\ref{constraint}) 
\begin{equation}
\frac{1}{\zeta
+b}=\frac{sM^{4}e^{-2\alpha\phi/M_{p}}+V_{1}}
{2\left[b\left(sM^{4}e^{-2\alpha\phi/M_{p}}+V_{1}(\upsilon
+\chi)\right)-V_{2}(\upsilon +\chi)\right]}
\label{zetavac}
\end{equation} 
one can check that in this case the dilaton and Higgs fields equations
(\ref{phief+ferm}) and (\ref{chief+ferm}) take the form of the canonical
scalar fields equations with the effective potential
\begin{equation}
V_{eff}(\phi ,\upsilon +\chi)=
\frac{\left[sM^{4}e^{-2\alpha\phi/M_{p}}+V_{1}(\upsilon
+\chi))\right]^{2}}
{4\left[b\left(sM^{4}e^{-2\alpha\phi/M_{p}}+V_{1}(\upsilon
+\chi)\right)-V_{2}(\upsilon +\chi)\right]}
\label{Veffvac}
\end{equation}

From this we immediately conclude that the stable vacuum of the
scalar fields ($<\phi>\equiv\overline{\phi}$ and $\upsilon$) is realized
as a manifold determined by the equation
\begin{equation}
 sM^{4}e^{-2\alpha\overline{\phi}/M_{p}}+V_{1}(\upsilon)=0
\label{vacscalars}
\end{equation}
provided that $V_{2}(\upsilon)<0$ in this degenerate vacuum. 
The masses of the dilaton and Higgs fields excitations above this
degenerate vacuum are respectively
\begin{equation}
m_{dilat}^{2}=\frac{\alpha^{2}M^{8}}{M_{p}^{2}|V_{2}(\upsilon)|}
e^{-4\alpha\overline{\phi}/M_{p}};\qquad
m_{higgs}^{2}=\frac{(V_{1}^{\prime}(\upsilon))^{2}}{|V_{2}(\upsilon)|}.
\label{masscalars}
\end{equation}
Notice that we did not assume any specific  properties of $V_{1}$  and
$V_{2}$ so far. If we wish to provide conditions for a big Higgs mass we
see from the second equation in (\ref{masscalars}) that there is no need
for big "pre-potentials"  $V_{1}(\upsilon)$  and  $V_{2}(\upsilon)$ but
rather they both can
be small as compared to a typical energy scale of particle physics,
however $V_{2}(\upsilon)$ must be very small.    

An  important feature of the degenerate  vacuum (\ref{vacscalars}) is that
the effective vacuum energy
density of the scalar fields is equal to zero without any sort of fine
tuning regardless of the detailed shape of the potentials $V_{1}$ and
$V_{2}$ as well as of the initial conditions. 
This fact has been very extensively
explored as a way to solve the cosmological constant
problem\cite{GK2,GK3}.    
In this paper, however, we will concentrate our 
attention on the applications of the theory to particle physics.

Notice that according to  Eq. (\ref{zetavac}), $\zeta =\infty$ in the 
degenerate vacuum (\ref{vacscalars}). However, in the presence of any 
small "contamination" by massive fermions, it follows from the constraint
(\ref{constraint}) that $\zeta$ is large but finite. Therefore we must
return to the general form of
the effective potential  (\ref{Veff}) which will  be small but non zero.
This means that zero vacuum energy is practically unachievable, and there
must be a correlation between the fermion content of the universe and the
vacuum energy. 

(ii) {\bf Case where fermion densities are of the typical laboratory
particle physics scales.} Assuming as it was done before that  
$M^{4}e^{-2\alpha\phi/M_{p}}$,
$V_{1}$  and $V_{2}$ are small as compared to the typical particle physics
energy densities of fermions\footnote{ In a recent
paper\cite{GK5} we studied  a simplified model without the Higgs field and
in
that case we had only to assume that fermion mass term
$m\overline{\Psi}^{\prime}\Psi^{\prime}$
is much larger than $M^{4}e^{-2\alpha\phi/M_{p}}$.}, we see from the
constraint
(\ref{constraint}) that now there
are no reasons for $\zeta$ to be large. On the contrary, it has to  be of
the same order as  the dimensionless parameters of the theory ($b$, $k$
and $h_{i}$) which we assume are of order one. So, for the case when
fermion densities are of the typical laboratory 
particle physics scales, $\zeta$ has to satisfy the simplified form of the
constraint(\ref{constraint}):  
\begin{equation}
(\zeta +b)[F_{E}(\zeta)\overline{E^{\prime}}E^{\prime} +
F_{U}\overline{U^{\prime}}U^{\prime}+F_{D}\overline{D^{\prime}}D^{\prime}]=0.
\label{decoupling}
\end{equation}
 To see the meaning of the
constraint in this case, let us take one single primordial fermionic
state:
or $E^{\prime}$, or $U^{\prime}$, or $D^{\prime}$.
Then we have three solutions for each of $\zeta^{(i)}$, \quad
($i=E^{\prime},U^{\prime},D^{\prime}$):  two constant
solutions are
defined by the condition $F_{i}(\zeta)=0$, i.e.
\begin{equation}
\zeta_{1,2}^{(i)}=\frac{1}{2}\left[k-3h_{i}\pm\sqrt{(k-3h_{i})^{2}+
8b(k-h_{i})
-4kh_{i}}\,\right], \qquad i=E^{\prime},U^{\prime},D^{\prime}
\label{zetapm}   
\end{equation}
and the third solution $\zeta +b=0$. 

The first two solutions correspond
to two different states of the $i's$ primordial fermion with {\bf 
different constant 
masses} determined by Eq.(\ref{muferm}) where we have to substitute
$\zeta_{1,2}^{(i)}$ instead of $\zeta$.
These two states can be identified with the first two generations of the 
physical leptons  and quarks.
Surprisingly (see also \cite{GK5}) that the same combination that we see
in the l.h.s. of
the constraint (\ref{decoupling}) appears in the last terms of
Eqs. (\ref{phief+ferm}) and (\ref{Tmn}) (we assume here that
$\zeta +b\neq 0$). Therefore, in the regime where the regular fermionic
matter (i.e. $u$ and $d$ quarks, $e^{-}$ and $\nu_{e}$) dominates, 
the last  terms of
Eqs. (\ref{phief+ferm}) and (\ref{Tmn}) {\it automatically} vanish.
In Eq. (\ref{phief+ferm}), this means that {\it the fermion densities
  are not a source for the dilaton
and thus the long-range forces disappear automatically}. Notice that
there is no need to require no interactions of the dilaton with fermionic
matter at all to have agreement with observations but  it is rather enough
that these interactions vanish in the appropriate regime where
regular fermionic
matter has the typical laboratory particle physics density.
In Eq. (\ref{Tmn}), the condition
(\ref{decoupling}) means that in the  region where the regular fermionic
matter dominates, the fermion energy-momentum   
tensor becomes equal to the canonical
energy-momentum tensor of fermion fields in GR
(see also Ref.\cite{GK5}).

The third solution $\zeta +b=0$ is singular one as we see from equations 
of motion. This means that one can not neglect the first two terms 
in the constraint (\ref{constraint}). Then 
instead of $\zeta +b=0$ we have to take the solution $\zeta +b\approx 0$
by solving $\zeta +b$ in terms of the dilaton and Higgs fields and the
primordial
fermionic fields themselves.
\begin{equation}
\frac{1}{\sqrt{\zeta_{3} +b}}\approx
\left[\frac{\upsilon \left[f_{E}(b-h_{E})\overline{E^{\prime}}E^{\prime}
+
f_{U}(b-h_{U})\overline{U^{\prime}}U^{\prime}
+f_{D}(b-h_{D})\overline{D^{\prime}}D^{\prime}\right]}
{4\sqrt{2}(b-k)\left[b\left(sM^{4}e^{-2\alpha\phi/M_{p}}+V_{1}\right)-V_{2}
\right]}\right]^{1/3}
\label{srtzeta}
\end{equation}

This leads to non-polynomial fermion interactions. A full 
treatment of the third family requires the study of quantum corrections 
and fermion condensates which will give the third family
appropriate masses.

{\bf V. DISCUSSION and CONCLUSIONS.} In this letter we have seen that a
generally covariant theory where we allow two measures of integration,
$\sqrt{-g}$ and the density $\Phi$ defined by Eq. (\ref{Phi}), yields
a natural geometric explanation of the origin of the fermion families in
particle
physics. Although we have seen this in the context of an $SU(2)\times
U(1)$ gauge theory, the conclusions remain unchanged if the $SU(3)$ color
gauge group is also added. 

Three fermion families are identified in the high fermion density
approximation, i.e. when energy densities of fermions are much bigger than
the scalar fields contributions to the constraint (\ref{constraint}).
Two of these fermion
families have constant particle masses (since $\zeta$  is a constant)
while for the third family a non-polynomial fermion interactions appear.
Interestingly enough that the effective coupling constants of these 
 non-polynomial interactions are dimensionless, which suggests that the
quantum  corrections of this theory may be meaningful.

This and other aspects of the quantization of this theory require further
study. Among the most interesting aspects of the quantum theory which
should be studied are the quantization of the measure fields
$\varphi_{a}$. In fact, we expect the "families birth effect" to be
closely
related to the functional integration over those measures fields. There 
we expect that functional integration will be restricted by the
configurations dictated by the constraint (\ref{constraint}). The
integration over the $\varphi_{a}$ fields should contain an integration
over $\zeta$ and integrations over volume preserving variables (i.e.
those that preserve the value of $\Phi$). At each point, the integration
over  $\zeta$ selects then the values where the constraint is satisfied
and for the fermion densities corresponding to laboratory conditions,
three possible values of  $\zeta$ are then selected. We hope to give more
details concerning these quantum aspects of the theory in a future
publication.

Finally, it is important to notice that the theory explained here allows
for transitions from a certain family to another.  One can indeed notice
from the constraint itself that distinct constant values
$\zeta^{i}_{1,2}$, Eq.(31),
corresponding to different quark families are obtained 
(again, for the quark densities corresponding to laboratory conditions)
only for pure states of primordial quarks (either $U^{\prime}$ or
$D^{\prime}$). For the case when the wave functions of
massive primordial quarks overlap at some space-time region, then, 
for very specific values of the ratios of the quarks densities some of the
three solutions of eq. (30) can coincide. We may call these points
"unification points". Since the distinction between families depends on 
our ability to differentiate between the three different solutions of
(30),  we see that such differentiation looses meaning at one of the
"unification points" where two, originally different values of $\zeta$,
describing two different families converge to the same value. At this
point the two families get "confused". Therefore, once one reaches some of
these unification points, it is clear that transitions from family to
family (or what is the same, the jump from one solution of eq. (30) to
another branch of eq. (30)) are possible. The
calculation of the amplitudes of these transitions appear to be
technically complicated but are in principle calculable. Therefore the
parameters of the Kobayashi-Maskawa mass matrix should be indeed
calculable as a function of the parameters of the theory.

In this paper we have ignored the question of a possible neutrino mass.
There is no problem to incorporate a neutrino mass in our formalism.
In fact, if we start with a single primordial neutrino field with a mass
term\footnote{it
does not matter how this is achieved: Majorana or Dirac (which
has right-handed neutrino and requires Higgs Yukawa coupling in the
$SU(2)\times U(1)$ gauge theory).},  we will  again   find that for pure
neutrino state in  laboratory conditions there are three possible values
of  $\zeta$, which give then three possible values for the neutrino mass,
i.e. different neutrino states. In this case the physics of neutrino 
mixing
will have some resemblance to the situation with quarks. It would be very 
important to see how the phenomenon of neutrino oscillations could appear
in the context of this theory.

As we have argued, the theory appears to provide a new way to address the
cosmological constant problem in  a manner similar to what we discussed in
Refs.\cite{GK2,GK3}. While the vacuum in the absence of massive fermions
is
at zero
cosmological constant, any fermion "contamination" does not allow the
vacuum with zero cosmological constant to be achieved. It appears then
that how much vacuum energy differs from zero, is correlated with how much
fermions are in the universe. This correlation might be a possible
mechanism for the explanation of the "cosmic coincidence"
problem\cite{St}.

Finally, we want to attract attention  to a very interesting effect:
when  densities of  the first two fermion families correspond to normal 
laboratory conditions, their interactions with the dilaton  disappear
automatically (see also \cite{GK5}).

{\bf  Acknowledgments.}
We are very grateful to A. Davidson and V.V. Kiselev for useful
discussions.

\end{document}